\documentclass[fleqn,usenatbib,nofootinbib]{mnras}
\usepackage[T1]{fontenc}
\usepackage{ae,aecompl}
\usepackage{booktabs}

\usepackage{graphicx}	
\usepackage{amsmath}	
\usepackage{amssymb}	
\usepackage{natbib}
\bibpunct{(}{)}{;}{a}{}{,} 
\usepackage{xcolor}
\usepackage[caption=false]{subfig}
\usepackage{etex}	
\usepackage{acronym}  
\usepackage{float}
\usepackage{makecell}
\usepackage{mathtools}
\usepackage{ulem}
\usepackage{footmisc}

\acrodef{BH}[BH]{black hole}
\acrodef{NS}[NS]{neutron star}
\acrodef{CO}[CO]{carbon-oxygen}
\acrodef{ZAMS}[ZAMS]{zero-age main sequence}

\newcommand\MCO{M_\mathrm{CO}}

\title[Compact remnant masses and kicks]{Simple recipes for compact remnant masses and natal kicks}

\author[Mandel \& M\"uller]{
            Ilya Mandel$^{1,2,3,4}$ and
            Bernhard M\"uller$^{1,2}$
\\
$^{1}$ School of Physics and Astronomy, Monash University, Clayton, Vic. 3800, Australia\\
$^{2}$ The ARC Centre of Excellence for Gravitational Wave Discovery -- OzGrav\\
$^{3}$ The ARC Centre of Excellence for All Sky Astrophysics in 3 Dimensions -- ASTRO 3D\\
$^{4}$ Birmingham Institute for Gravitational Wave Astronomy and School of Physics and Astronomy,\\ University of Birmingham,  B15 2TT, Birmingham, UK
}

\date{Accepted 2020 September 28. Received 2020 September 25; in original form 2020 June 15.}
\pubyear{2020}

\begin{document}

\label{firstpage}
\pagerange{\pageref{firstpage}--\pageref{lastpage}}
\maketitle

\begin{abstract}
Based on recent results from three-dimensional supernova simulations and semi-analytical parametrised models, we develop analytical prescriptions for the dependence of the mass of neutron stars and black holes and the natal kicks, if any, on the pre-supernova carbon-oxygen core and helium shell masses.  Our recipes are probabilistic rather than deterministic in order to account for the intrinsic stochasticity of stellar evolution and supernovae.  We anticipate that these recipes will be particularly useful for rapid population synthesis, and we illustrate their application to distributions of remnant masses and kicks for a population of single stars. 
\end{abstract}

\begin{keywords}
supernovae: general -- stars: neutron -- stars: black holes
\end{keywords}

\section{Introduction}

Stars with initial masses above $\gtrsim 8 M_\odot$ end their lives as compact objects: neutron stars or black holes.  The amount of mass ejected during the supernova explosion, which may accompany the collapse of the stellar core, determines the remnant mass.  Meanwhile, an asymmetry in the explosion leads to a natal kick of the remnant.  If the massive star was a component of a stellar binary, as most massive stars are \citep{Sana:2012}, the mass loss and natal kick determine the future evolution of the system.  Will it be disrupted by the explosion? Will it become an X-ray binary?  Could it merge as a gravitational-wave source?  Even for single stars, natal kicks play critical roles; for example, they determine the fraction of compact objects retained in globular clusters.  Rapid stellar and binary population synthesis codes rely on recipes for predicting remnant masses and kicks from pre-supernova properties.  


Detailed modelling of core-collapse supernovae of evolved massive stars is a long-standing problem in computational astrophysics \citep[see, e.g.][for reviews]{Janka:2012,Burrows:2013,Mueller:2020}.  Below, we touch on some of the challenges associated with ab-initio modelling. 

On the other hand, there are no direct observations that could tie a pre-supernova stellar mass to the remnant mass or kick for an individual star, although a few supernova remnants such as Cas A allow for tentative remnant mass estimates with some information on the ejecta mass coming from light echoes \citep[e.g.,][]{Hwang:2012}, and a few tens of supernova progenitors have been directly observed \citep{Smartt:2015,VanDyk:2017}. 

Consequently, remnant mass and kick predictions are typically based on either simple analytical models, extrapolations of numerical simulations, or on observational constraints on entire populations of stars.  \citet{Hurley:2000,Fryer:2012} provide some the commonly used prescriptions for remnant masses, with more recent models available from, e.g., \citet{Mueller:2016} and \citet{PattonSukhbold:2020}.  Remnant kick prescriptions for neutron stars are typically based on observed velocities of single pulsars \citep{LyneLorimer:1994,HansenPhinney:1997,CordesChernoff:1998,Arzoumanian:2002,Hobbs:2005,FGKaspi:2006,IgoshevVerbunt:2017}.  However, using these prescriptions directly entails applying the same kick distributions to all neutron stars, regardless of pre-supernova properties,  perhaps with scaled kicks for black holes \citep[e.g.,][]{Fryer:2012,Belczynski:2012}.  Some population synthesis codes develop their own prescriptions, e.g., \citet{Belczynski:2008} or \citet{Spera:2015} for remnant masses and \citet{VignaGomez:2018} or \citet{BrayEldridge:2016,BrayEldridge:2018,GiacobboMapelli:2020} for natal kicks.

Here, we use recent findings from three-dimensional (3D) supernova simulations and parametrised one-dimensional (1D) and one-zone models 
as a starting point to predict the remnant mass from the progenitor core mass.  Unlike other models (see \citealt{Clausen:2015} for a notable exception), we do not attempt to deterministically prescribe a unique remnant mass, but rather account for the stochasticity in the pre-supernova stellar evolution and the stellar collapse itself by providing a probabilistic recipe.  We use the same models to predict natal kicks, which are thus coupled to the explosion properties and the remnant masses.  Our recipes have several tuneable parameters that are only approximately predicted by the numerical models.  We propose that these should be calibrated by observations.  We discuss the insights gained from supernova models in Section \ref{sec:models}; introduce the parametrised recipes directly usable in population synthesis in Section \ref{sec:recipes}; illustrate the consequences of these recipes for single stellar evolution in Section \ref{sec:single}; and briefly summarise in Section \ref{sec:conclusion}.

\section{Physical model}\label{sec:models}

\subsection{Phenomenology of 1D and 3D supernova simulations}
\label{sec:model_phenomenology}

In recent years, parametrised models and first-principle
3D simulations have made significant progress in
establishing the relationship between the supernova progenitor,
the explosion, and the properties of the remnant in the neutrino-driven supernova
paradigm. Various studies have addressed the ``explodability'' of supernova progenitors,
using 1D simulations and semi-analytic models
\citep[e.g.,][]{Oconnor:2011,Ugliano:2012,Pejcha:2015,Ertl:2016b,Mueller:2016,Ebinger:2019,Ertl:2020}. Despite variations
in detail, the findings suggest that explodability
can be gauged by a few key stellar structure parameters.
Physically, the key determinants that favor
a successful explosion are a low Fe-Si core
mass and a low density in the oxygen (O) shell
\citep{Ertl:2016b}, although other metrics
for explodability like
the more familiar compactness parameter
\citep{Oconnor:2011} or the binding energy of
the shells outside the Fe-Si core are strongly
correlated with these parameters. These structural
properties of the core region depend non-monotonically
on \ac{ZAMS} mass (for single stars), He core mass, and \ac{CO}
core mass $M_\mathrm{CO}$, leaving ``islands of explodability'' among
rather massive progenitors. Nonetheless, from
a coarse-grain point of view, there \emph{is}
a trend of decreasing explodability with increasing
He and CO core mass.

First-principle 3D simulations are not yet capable
of surveying stellar explodability in a similar manner
as  parametrised supernova explosion models. There
are tentative indications, however, that neutrino-heating
can drive shock expansion in progenitors with cores
of high mass and compactness
\citep{Chan:2018,Ott:2018,Burrows:2020,Powell:2020},
which is seemingly in contrast to the previously mentioned findings.
These results are not likely to fundamentally overturn
the trend toward lower explodability with higher
He or CO core mass, since the incipient explosions
in progenitors with high-mass cores may be stifled
on longer time scales as the shock propagates through
the tightly bound massive envelope. If shock revival
works effectively in progenitors with high-mass cores,
the net effect may merely be to introduce a transition
regime of successful explosions with considerable
fallback and black-hole formation between the regime
of neutron star formation (for less massive cores)
and quiet black hole formation (for the most massive cores).
This would be compatible with constraints
on the progenitors of observed transients from
pre-explosion images, according to which explosions
from high-mass progenitors are rare
\citep{Smartt:2009,Smartt:2015}, as well as with the apparent disappearance of massive stars without an explosion \citep{Adams:2017}.

Some systematic dependencies of the compact remnant properties 
can already be discerned from paramaterised and first-principle
supernova models.  In the case of successful
explosions with neutron star formation, there is a rather robust
correlation between the progenitor's Fe-Si core mass and
the neutron star mass $M_\mathrm{NS}$
\citep[e.g.,][]{Ertl:2016b,Sukhbold:2016,Mueller:2016}, because
neutrino-driven explosions are often triggered by the drop
in mass accretion rate when the collapsing O shell reaches the stalled shock. 
The dependence of $M_\mathrm{NS}$ on the He and \ac{CO} core mass
is not monotonic, but there is a loose positive correlation
with significant scatter due to the intricacies of the late
stellar  evolution stages and of the neutrino-driven mechanism.
A linear dependence of $M_\mathrm{NS}$ on $M_\mathrm{CO}$ with broad scatter would thus seem the most natural first-order approximation.

In case the shock is successfully revived, the explosion energy 
correlates loosely with core mass and compactness, and with
the neutron star mass
\citep{Nakamura:2015,Mueller:2016,Mueller:2019}
because
the neutron star mass (via the neutron star surface
temperature) and the accretion rate determine neutrino emission
and neutrino heating. There is also a loose correlation between \ac{NS} mass, explosion energy, and NS kick velocity since the explosion
energy sets the scale for the momentum asymmetry that
can be achieved in the innermost ejecta, even though
the degree of asymmetry is expected to exhibit stochastic variations
\citep{Janka:2017,VignaGomez:2018}. 3D simulations indeed
indicate substantial scatter in the kicks \citep{Mueller:2019,Powell:2020},
especially for progenitors with low-mass cores
\citep{Gessner:2018,Stockinger:2020}, see
e.g., Figure~12 in \citet{Mueller:2019}.

Due to large-scale lepton fraction asymmetries in
the proto-neutron star convection zone,
anisotropic neutrino emission can also impart
kicks of a few tens of $\mathrm{km}\, \mathrm{s}^{-1}$
onto the neutron star even without
pronounced asymmetries in the ejecta \citep{Stockinger:2020}.

In the case of black hole formation, the systematics of the remnant
properties are less securely understood, partly because
there are different scenarios for black hole formation.
When a black hole is formed because the shock is never revived,
one does not expect any kick, but if a hydrogen envelope
is present it may (partly) become unbound as the emission
of neutrinos during the proto-neutron star phase decreases
the gravitational mass of the star 
\citep{Nadyozhin:80,Lovegrove:2013,Coughlin:2018}. In this
case, one can identify black hole mass with the helium core mass
as a first-order estimate.

In other cases, a black hole may be formed by fallback after
successful shock revival. Fallback can occur via the deceleration
of the inner ejecta by the reverse shock \citep{Chevalier:1989},
but this mechanism usually only adds $\lesssim 10^{-2} M_\odot$
onto the remnant \citep{Ertl:2016}. Substantial fallback
can occur, however, if ongoing accretion 
after shock revival pushes the neutron star to collapse,
in which case the dynamics of the fallback is inherently multi-dimensional \citep{Chan:2018,Chan:2020}. Based on the work of \citet{Chan:2020},
one can distinguish two limiting scenarios in this regime
of strong fallback. 

If the explosion energy is sufficiently high,
then the post-shock velocity will eventually reach escape velocity
even though the explosion energy is continuously decreased as
the shock sweeps up bound matter. Up to this point, a substantial
fraction of the shocked matter will be channelled around
the expanding neutrino-heated bubbles and accreted onto the
black hole; afterwards shocked material will become unbound
and be ejected. Since the expanding bubbles that eventually
escape usually develop a unipolar or bipolar structure early
in the explosion, they will carry a significant net momentum at
the point when accretion freezes out, and the black hole, must
by momentum conservation receive a substantial kick that
can reach several $100 \, \mathrm{km}\, \mathrm{s}^{-1}$.

If, on the other hand, the Mach number of the shock comes close
to unity as the accumulation of bound matter drains the
explosion energy, the blast wave will transition to the weak
shock regime, i.e., the shock will propagate as a sonic
pulse whose energy is approximately conserved \citep{Mihalas:1984}.
As the shock approaches the stellar surface, it will
eventually leave the weak shock regime again and shed a
part of the envelope. The final mass cut can be estimated
by considering when the post-shock velocity eventually
matches the escape velocity. The sonic pulse is also spherical
in this case, and hence the ejecta momentum and the black
hole kick will be small, though residual kicks 
of order $\mathcal{O}(10\, \mathrm{km}\, \mathrm{s}^{-1})$ are still possible
\citep{Chan:2018}.

The transition from the regime of weak
fallback to strong and ultimately complete fallback is
governed by the ratio of the envelope binding energy
to the initial explosion energy. Since the
explosion energy is limited to $\mathord{\sim}2\times 10^{51}\, \mathrm{erg}$ in the neutrino-driven paradigm, one would expect that the envelope binding energy (which is correlated
with the mass of the oxygen, neon and carbon shells) is
the decisive parameter that dictates the amount of fallback.

In reality, fallback is a more complicated,
genuinely multi-dimensional hydrodynamic process.
Introducing two discrete scenarios with sharp 
``switches'' for terminating fallback allows us to 
obtain qualitatively correct fallback masses and 
capture the dependence of fallback on stellar 
parameters, but likely introduces an overly ``clumpy'' distribution of remnant masses
(as shown in Figure~\ref{fig:masses})
relative to what may be expected in nature. There are as yet
too few multi-dimensional models available to sample
the distribution of black hole masses in fallback
supernovae, but the more gradual termination 
of accretion seen in extant multi-D models
\citep{Chan:2018,Chan:2020} will likely
result in a smoother distribution, and probably
a bigger stochastic scatter than predicted by
our semi-analytic model. Even though they
cannot capture the nuances of multi-dimensional fallback processes, hydrodynamic
simulations in spherical symmetry \citep{Ertl:2020}
already suggest a slightly broader distribution
of black hole masses that reaches down to
the most massive neutron star masses and does
not exhibit any mass gap.

\subsection{Updated semi-analytic supernova model}\label{sec:analytics}
As a basis for more a more agnostic parametrisation of
the relation between stellar progenitor parameters
and the compact remnant mass and kick, which will
be discussed in Section~\ref{sec:recipes}, we use
results from the semi-analytic supernova model of
\citet{Mueller:2016}. This model reproduces
much of the phenomenology revealed by the more detailed 
multi-dimensional simulations.  Meanwhile, its parameters
have been chosen to comply with observational
constraints (e.g., on supernova explosion energies)
that multi-dimensional simulations do not yet comply 
with. In particular, the model shows the same loose
correlation between progenitor compactness, explosion energy, and the neutron star mass and kick that
has been alluded to in our discussion of multi-dimensional simulation in Section~\ref{sec:model_phenomenology}
\citep{Nakamura:2015,Mueller:2019,Burrows:2020}.

While this model has already
been used to formulate fits for
neutron star masses and kicks in \citet{VignaGomez:2018}, the 
original model of \citet{Mueller:2016} only included
a crude all-or-nothing treatment of fallback, and therefore
does not provide useful predictions for black hole masses
and  kicks. We have therefore updated the semi-analytic
model to reflect some of the phenomenology of 
multi-dimensional simulations of black hole formation in fallback supernovae.

During the pre-explosion phase, the updated semi-analytic model follows
the original prescriptions of \citet{Mueller:2016}  to estimate
whether and when shock revival occurs based on scaling laws for the
neutrino emission, the quasi-stationary structure of the supernova
core, and the neutrino heating conditions. However,
we use a slightly smaller value $\zeta=0.7$ for the
accretion efficiency for better agreement with
the observational constraints on supernova progenitor
masses \citep{Smartt:2015}.
The treatment of
the explosion phase has been modified slightly. After shock revival
at the initial mass cut $M_\mathrm{i}$,
the original model computed the explosion energy as a function
of mass coordinate by taking into account neutrino heating,
nuclear energy release, and the accumulation of bound material
by the shock. The final mass cut $M_\mathrm{f}$ was determined by the 
condition
that the shock velocity  $v_\mathrm{sh}$ (estimated from the explosion
energy and ejecta mass following \citealt{Matzner:1994}) exceeded
the escape velocity $v_\mathrm{esc}$.  If this condition was not met before the
gravitational neutron star mass 
$M_\mathrm{grav}$ reached $M_\mathrm{max}=2.05 M_\odot$, or if
the explosion energy dropped to zero, the model assumed that
the entire star collapses to a black hole. We have modified this procedure as follows:
\begin{enumerate}
    \item Once $M_\mathrm{grav}=M_\mathrm{max}$, we switch off
    further energy input by neutrinos, but allow the shock to propagate
    further after black hole formation, and assume that the explosion energy
    changes according to the release of nuclear energy by explosive burning
    and the accumulation of bound matter following Equation~(49) of
    \citet{Mueller:2016}.
    \item If $v_\mathrm{sh}=v_\mathrm{esc}$ is
    reached after black hole formation, we assume that accretion stops.
    The corresponding mass coordinate defines the final mass cut
    mass $M_\mathrm{f}$. The mass $M_\mathrm{asym}=
    M_\mathrm{f}-M_\mathrm{i}$ of the strongly asymmetric inner ejecta
   determines the kick. The explosion energy is still evolved
    further until shock breakout, or until one of the following conditions
    is met. This channel corresponds to the scenario
    of black hole formation with weak fallback in \citet{Chan:2020}.
    \item If $v_\mathrm{sh}$
    drops below the pre-shock sound speed $c_\mathrm{s}$ (which is 
    smaller than the escape velocity), we assume a
    transition to the weak shock regime. In this case, the
    (small) explosion energy at the transition point is conserved
    as the shock travels to the surface as a sound pulse, 
    and the final mass cut $M_\mathrm{f}$ is  determined as the mass
    coordinate where $v_\mathrm{sh}>v_\mathrm{esc}$ somewhere
    further outside. This channel corresponds to the strong
    fallback scenario in \citet{Chan:2020}. In our supernova
    model we do not calculate any kick for this channel for want
    of a suitable analytic theory. The results of \citet{Chan:2018,Chan:2020}
    suggest, however, that black hole kicks of several $\times 10 \, \mathrm{km}\, \mathrm{s}^{-1}$ will be reached.
    \item Following \citet{Lovegrove:2013}, we assume that
    the hydrogen shell (if present) is always ejected, and cap
    the black hole mass at the helium core mass $M_\mathrm{He}$.
\end{enumerate}
Both for neutron stars and black holes, we estimate the typical
value of the kick from $M_\mathrm{f}$ and $M_\mathrm{i}$
as in \citet{VignaGomez:2018},
\begin{equation}
\label{eq:kick}
v_\mathrm{kick}=
0.16\frac{\sqrt{E_\mathrm{expl}(M_\mathrm{f}-M_\mathrm{i})}}{M_\mathrm{grav}},
\end{equation}
where $E_\mathrm{expl}$ is the explosion energy.
Since the mass $M_\mathrm{f}-M_\mathrm{i}$ of the asymmetric inner ejecta is tightly correlated
with the explosion energy 
 $E_\mathrm{expl}$, this is effectively tantamount to $v_\mathrm{kick}
 \propto  E_\mathrm{expl}$ in the case of neutron-star forming
 progenitors with little fallback, though  Equation~(\ref{eq:kick}) makes
 the connection to momentum conservation more transparent.
 Such a scaling of the kick velocity with explosion energy (with
 some scatter) has
 indeed been found in self-consistent 3D simulations, see, e.g., Figure~12 in \citealp{Mueller:2019}.
 
To obtain the remnant gravitational mass $M_\mathrm{grav}$, we subtract
the neutron star binding energy from $M_\mathrm{f}$, but no such
deduction is made in the case of black hole formation.
The stochastic distribution of the kick velocity is not specified
any further in the semi-analytic model, and will be dealt
with by a parametrised prescription in Section~\ref{sec:recipes}.
The kick estimate from Equation~(\ref{eq:kick}) is best
understood as the mode of a broad, probably top-heavy distribution,
whose precise shape cannot be predicted based on first principles
and a small corpus of 3D simulations yet.

In practice, $E_\mathrm{expl}$ and
the mass $M_\mathrm{f}-M_\mathrm{i}$ of the asymmetric
inner ejecta are tightly  correlated. Because
of the correlation of $E_\mathrm{expl}$ with
$M_\mathrm{CO}$, this suggests a linear dependence of
the kick velocity on $M_\mathrm{CO}$ as a first-order
approximation. For low-mass iron core supernovae and
electron capture supernovae, where $M_\mathrm{NS}\approx
M_\mathrm{CO}$, 3D simulations show very small kicks
\citep{Mueller:2018,Stockinger:2020}, and hence
it is natural to anchor this approximate linear
dependence at zero, i.e., 
$v_\mathrm{kick}\propto \sqrt{E_\mathrm{expl}(M_\mathrm{f}-M_\mathrm{i})}
\propto (M_\mathrm{CO}-M_\mathrm{NS})$.

We stress that this procedure for estimating fallback after
black hole formation rests on a crude simplification of the complex dynamics
in multi-dimensional simulations. It does nonetheless qualitatively
reproduce the salient features of these simulations; low-mass
black holes made in relatively powerful explosion will get sizeable
kicks, whereas high-mass black holes formed in 
weak explosions (which tend to have 
strong fallback) will not.
Unlike the simpler prescriptions for fallback
and black hole kicks
\citep{Fryer:2012,Spera:2015,BrayEldridge:2016,BrayEldridge:2018,Kruckow:2018,GiacobboMapelli:2020}, it is
the detailed structure of the innermost shells of the progenitor that determines whether an explosion enters
the regime of weak or strong fallback. As a result,
our semi-analytic explosion model predicts that in the
channel for \ac{BH} formation, weak and strong fallback coexist over a wide range of \ac{CO} core masses, and even
coexist with the neutron star formation channel.
This behaviour is genuinely different from 
single-valued prescriptions for the remnant properties
and significantly changes the mapping from
$M_\mathrm{CO}$ to  the remnant mass (and hence also
the remnant kick). The distinct remnant mass ``branches''
in Figure~\ref{fig:masses} should be
compared, e.g., to Figure~7 in \citet{Spera:2015}
or Figure~5 in \citet{Fryer:2012}.  This more complex dependence of
the remnant properties on  $M_\mathrm{CO}$
emerges naturally from a more detailed treatment
of the supernova explosion physics, not only
in our semi-analytic model but also in other
studies \citep[e.g.,][]{Ugliano:2012,Pejcha:2015,Sukhbold:2016,Ertl:2020}.
It stands to reason that this complexity will be reflected
in the range and distribution of compact remnants and compact binary
systems that can be realised in nature, and therefore 
ought to be taken into account in population synthesis codes.

\section{Parametrised prescription}\label{sec:recipes}

Our next goal is to provide a simple prescription with a small number of parameters for the remnant mass and kick as a function of the \ac{CO} core and He shell mass of the progenitor (we assume that any hydrogen envelope will always be unbound during the supernova \citep{Lovegrove:2013,Coughlin:2018}, if not previously stripped by winds or binary interaction).   While we suggest likely values for these parameters based on our semi-analytic model, the exact values should be updated by comparisons with observations. 

We assume that the combined process of stellar evolution and supernova explosion and collapse introduces genuine stochasticity on top of deterministic dependencies on stellar parameters.  Therefore, unlike all previous recipes, our model is probabilistic in its predictions.

\begin{table*}
\centering
\begin{tabular}{|l|l|l|}
\hline
Parameter  & Default value  & Meaning \\
\hline
\multicolumn{3}{|c|}{Mass ranges}\\
{$\boldsymbol{M_1}$} & $2.0 M_\odot$ & Max CO core mass leading to 100\% NS\\
{$\boldsymbol{M_2}$} & $3.0 M_\odot$ & Break in NS mass distribution fits\\
{$\boldsymbol{M_3}$} & $7.0 M_\odot$ & Min CO core mass leading to 100\% BH\\
{$\boldsymbol{M_4}$} & $8.0 M_\odot$ & Min CO core mass leading to 100\% fallback\\
\hline
\multicolumn{3}{|c|}{Remnant mass parameters}\\
$\mu_1$ & $1.2 M_\odot$ & mean NS mass for $\MCO < M_1$\\
$\sigma_1$ & $0.02 M_\odot$ & NS mass scatter for $\MCO < M_1$\\
$\mu_{2a}$ & $1.4 M_\odot$ & NS mass offset for $M_1 \leq \MCO < M_2$\\
$\mu_{2b}$ & $0.5$ & NS mass scaling for $M_1 \leq \MCO < M_2$\\
$\sigma_2$ & $0.05 M_\odot$ & NS mass scatter for $M_1 \leq \MCO < M_2$\\
$\mu_{3a}$ & $1.4 M_\odot$ & NS mass offset for $M_2 \leq \MCO < M_3$\\
$\mu_{3b}$ & $0.4$ & NS mass scaling for $M_2 \leq \MCO < M_3$\\
$\sigma_3$ & $0.05 M_\odot$ & NS mass scatter for $M_2 \leq \MCO < M_3$\\
$\mu_\mathrm{BH}$ &  $0.8$ & BH mass scaling for $M_1 \leq \MCO < M_4$\\
$\sigma_\mathrm{BH}$ & $0.5 M_\odot$ & BH mass scatter for $M_1 \leq \MCO < M_4$\\
$M_\mathrm{NS, min}$ & $1.13 M_\odot$ & minimal NS mass from core-collapse SN\\
{$\boldsymbol{M_\mathrm{NS, max}}$} & $2.0 M_\odot$ & maximal NS mass\\
\hline
\multicolumn{3}{|c|}{Natal kick parameters}\\
{$\boldsymbol{v_\mathrm{NS}}$} & 400 km s$^{-1}$ & NS kick scaling prefactor\\
{$\boldsymbol{v_\mathrm{BH}}$} & 200 km s$^{-1}$ & BH kick scaling prefactor\\
$\sigma_\mathrm{kick}$ & $0.3$ & fraction kick scatter\\
\hline
\end{tabular}
\caption{The list of parameters describing our remnant mass and kick model.  The most uncertain parameters that can benefit from observational constraints appear in bold.}\label{table:params}
\end{table*}

The key parameters governing the mass and kick, along with their default values, are identified in Table~\ref{table:params}.  The remnant mass prescription depends predominantly on the \ac{CO} core mass, $\MCO$.  We split the entire mass range into 5 sub-ranges with boundaries $M_1, M_2, M_3, M_4$ depending on the value of $\MCO$.

First, we determine the remnant type.  If $\MCO < M_1$, the remnant is always a \ac{NS}.    If $M_1 \leq \MCO < M_3$, the remnant is \ac{BH} with probability $p_\mathrm{BH} = (\MCO - M_1)/(M_3-M_1)$ and a \ac{NS} with probability $p_\mathrm{NS} = 1-p_\mathrm{BH}$.  If $\MCO \geq M_3$, the remnant is always a \ac{BH}.

If the remnant is a \ac{BH}, it has a probability $p_\mathrm{cf}$ of being formed by complete fallback, in which case the remnant mass is equal to the total He core mass (\ac{CO} core mass combined with He shell mass). The complete fallback probability is $p_\mathrm{cf}=1$ if $\MCO \geq M_4$.  Otherwise, for $M_1 \leq \MCO \leq M_4$, if the remnant is a black hole, its complete fallback probability is $p_\mathrm{cf} = (\MCO - M_1)/(M_4-M_1)$. 
The increasing probability of complete fallback with increasing $\MCO$ is compatible with a higher prevalence of this scenario at high $\MCO \gtrsim 7 M_\odot$ in the semi-analytic model (Figure~\ref{fig:masses}).

If the remnant is a \ac{BH} but is not formed by complete fallback, the remnant mass follows a normal distribution with mean $\mu_\mathrm{BH} \MCO$ and standard deviation $\sigma_\mathrm{BH}$.

Inspired by the parametrised semi-analytic model, we introduce several branches for the \ac{NS} mass.  If the remnant is a \ac{NS}, its mass is given by a normal distribution $N(\mu, \sigma^2)$ where the mean $\mu$ and standard deviation $\sigma$ are:
\begin{itemize} 
\item If $\MCO < M_1$, then $\mu = \mu_1$ and $\sigma = \sigma_1$.
\item  If $M_1 \leq \MCO < M_2$, then $\mu = \mu_{2a} + \mu_{2b} (\MCO - M_1)/(M_2-M_1)$ and $\sigma = \sigma_2$.
\item  If $M_2 \leq \MCO < M_3$, then $\mu = \mu_{3a} + \mu_{3b} (\MCO - M_2)/(M_3-M_2)$ and $\sigma = \sigma_3$.
\end{itemize}

We assume that \acp{NS} have a minimum mass of $M_\mathrm{NS, min}$ and a maximum mass of $M_\mathrm{NS, max}$, and therefore re-draw any remnant mass guesses that fall outside this allowed range (and similarly for any \ac{BH} mass guesses below $M_\mathrm{NS, max}$).

The remnant mass prescription above is already a recipe for the gravitational mass of neutron stars.
For black holes, we expect that only $0.1 M_\odot c^2 \sim \mathrm{few} \times 10^{53}$ erg of energy can be lost in neutrinos during the collapse.  This follows from the assumption that the total post-bounce neutrino luminosity is of order $10^{53}$ erg s$^{-1}$ and the \ac{BH} formation occurs on a timescale of a few seconds \citep{Mirizzi:2016,Chan:2018}, which is substantially shorter than the Kelvin-Helmholtz
timescale of the proto-neutron star.  Therefore, all final \ac{BH} remnant masses should be reduced by at most $0.1 M_\odot$ from the estimates given above, a correction which we ignore given the other uncertainties present (cf.~the much larger 10\% mass loss assumed by  \citet{Fryer:2012}).   

Although all simulated stars were single stars, we assume that binary interactions do not significantly impact the core structure, and therefore all prescriptions can be carried over directly to stars in interacting binaries, including those whose helium envelopes had been stripped by previous interactions (e.g., case BB mass transfer leading to ultra-stripped supernovae, \citealt{Tauris:2015}).  

We follow the prescription of \citet{Hurley:2000} with partial amendments from \citet{Belczynski:2008} and \citet{Fryer:2012} to determine which stars undergo electron-capture and regular core-collapse supernovae.  Stars with core masses above $2.25 M_\odot$ at the base of the asymptotic giant branch undergo core-collapse supernovae once their carbon-oxygen core reaches the mass threshold of Eq.~(75) of \citet{Hurley:2000}, where we use $1.38 M_\odot$ in lieu of the Chandrasekhar mass.  The remnant masses are then computed as described above.  If the helium core mass is below $2.25 M_\odot$ but above $1.6 M_\odot$ at the base of the asymptotic giant branch and $\MCO$ reaches $1.38 M_\odot$ during subsequent evolution, the star is assumed to form a \ac{NS} with a gravitational mass of $1.26 M_\odot$ in an electron-capture supernova.

Our models suggest that the inner ejecta during supernovae leading to \ac{NS} formation, which are coupled to the remnant and can therefore provide an asymmetric natal kick, have a similar degree of asymmetry and a similar velocity across all simulations.  Outer ejecta, including most of the helium envelope, are expelled before significant asymmetry is built up and hence do not provide a natal kick, while any material that falls back cannot provide a kick. 
Conservation of momentum implies that $M_\mathrm{NS} v_\mathrm{kick} = \alpha (M_\mathrm{CO}-M_\mathrm{NS}) v_\mathrm{ej}$, where $\alpha$ is the asymmetry parameter, $v_\mathrm{ej}$ is the typical ejecta velocity, and $M_\mathrm{CO}-M_\mathrm{NS}$ is a proxy for the mass of the material ejected asymmetrically while there is significant coupling with the remnant (see Section \ref{sec:analytics}).
Therefore, our model for the mean natal kick for neutron stars is
\begin{equation}\label{eq:NSkick}
\mu_\mathrm{kick} = v_\mathrm{NS} \frac{\MCO-M_\mathrm{NS}}{M_\mathrm{NS}}\ .
\end{equation}
We note that the physical reason for this scaling lies in the dependence of the explosion energy on
CO core mass, but for the purpose of constructing a parametrised recipe, it is immaterial whether there is a direct or and indirect dependence on $M_\mathrm{CO}$. We allow for stochasticity by adding an additional component drawn from a Gaussian distribution with standard deviation $\sigma_\mathrm{kick} \mu_\mathrm{kick}$.  

\citet{BrayEldridge:2016,BrayEldridge:2018,GiacobboMapelli:2020} previously proposed the use of momentum-preserving natal kick recipes in population synthesis studies.  These recipes have a generally similar form to that of Eq.~(\ref{eq:NSkick}), but with one significant distinction: they use the ratio of the ejecta mass to the remnant mass to determine the kick, while we use the ratio of the ejected carbon-oxygen core mass to the remnant. Our models indicate that this is a better proxy for the portion of the ejecta that are coupled to the asymmetry which builds up during the later stages of the collapse. Estimating the kick based on the total
ejecta mass would suggest substantially larger kicks for
non-stripped supernova progenitors, which is sharply contradicted
by 3D simulations \citet{Mueller:2019}. Moreover, \citet{BrayEldridge:2016,BrayEldridge:2018} fits include an offset to the scaling with this ratio, and in the latter case allow for negative kick velocities, i.e., remnant velocities in the direction in which greater momentum is carried away. While \citet{BrayEldridge:2018} explain this as being sourced by the gravitational tugboat mechanism \citep{Janka:2017}, this appears to unphysically contradict conservation of momentum.
Finally, we consider stochasticity in the kicks to be an intrinsic feature of stellar evolution and supernova instabilities, and our kick models reflect this along with the remnant mass models.

We do not use a distinct prescription to provide reduced natal kicks for electron-capture or ultra-stripped supernovae, as the low \ac{CO} core masses of their progenitors already provide the necessary kick reduction \citep{Tauris:2015,BeniaminiPiran:2016,VignaGomez:2018}.

Observations of black-hole X-ray binaries and micro-lensed black holes indicate that relatively low-mass black holes may experience kicks of a few tens to $\sim100$ km s$^{-1}$ \citep{Willems:2005,Fragos:2009,Mandel:2015kicks,Mirabel:2016,WyrzykowskiMandel:2019,Atri:2019} and possibly even a few hundred km s$^{-1}$ \citep{Repetto:2017}.  
Motivated by the discussion in Section \ref{sec:models},
we use the same functional form for black hole kicks as for neutron star kicks (Eq.~(\ref{eq:NSkick})), but with a reduced prefactor $v_\mathrm{BH}$ instead of $v_\mathrm{NS}$:
\begin{equation}\label{eq:BHkick}
\mu_\mathrm{kick} = v_\mathrm{BH} \frac{\max({\MCO-M_\mathrm{BH},\ 0)}}{M_\mathrm{BH}}\ .
\end{equation}
When a black hole is formed through complete fallback, $\MCO-M_\mathrm{BH} \leq 0$, so such black holes do not experience a natal kick in our model.

Remnant kick magnitudes are constrained to be positive and are re-drawn if the initial guess is negative.


\section{Single-stellar evolution}\label{sec:single}

\begin{figure}
\centering
\includegraphics[width=\columnwidth]{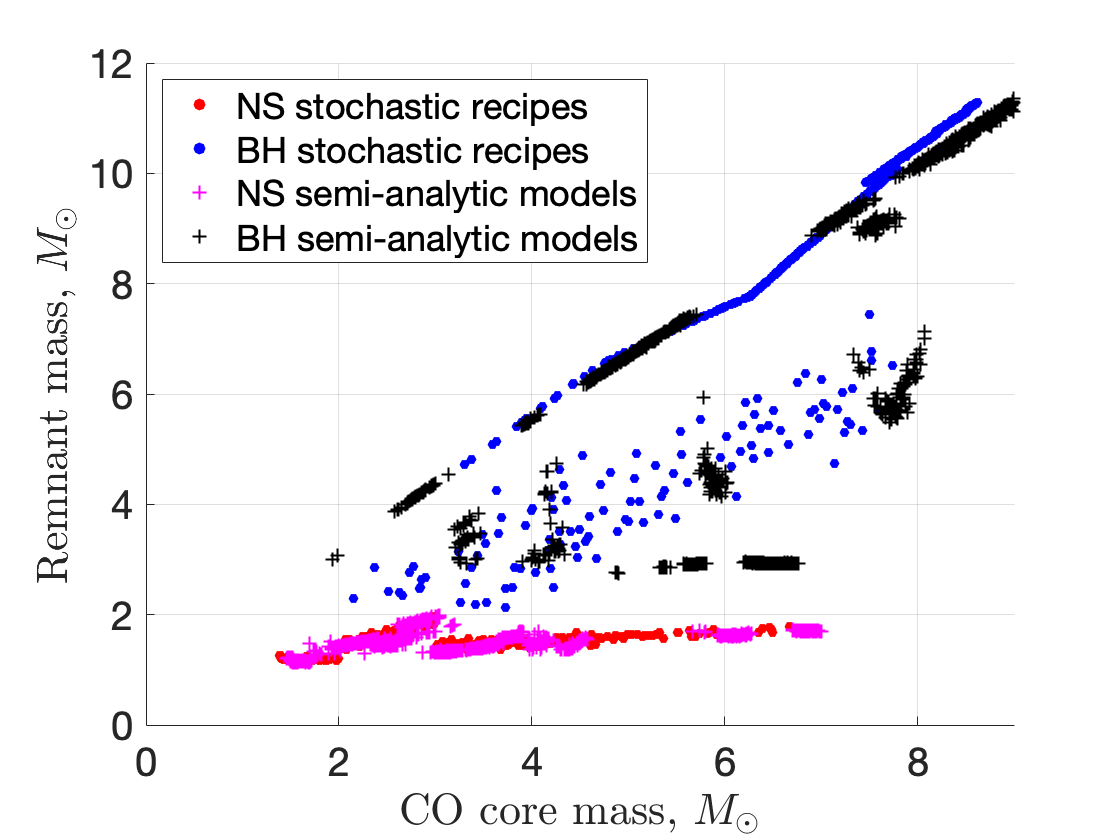}
   \caption{Remnant masses as a function of CO core masses from single-star simulations: a comparison between detailed stellar evolution coupled with the semi-analytic supernova model against COMPAS stellar evolution and our stochastic recipes.} 
    \label{fig:masses}
\end{figure}

\begin{figure}
\centering
\includegraphics[width=\columnwidth]{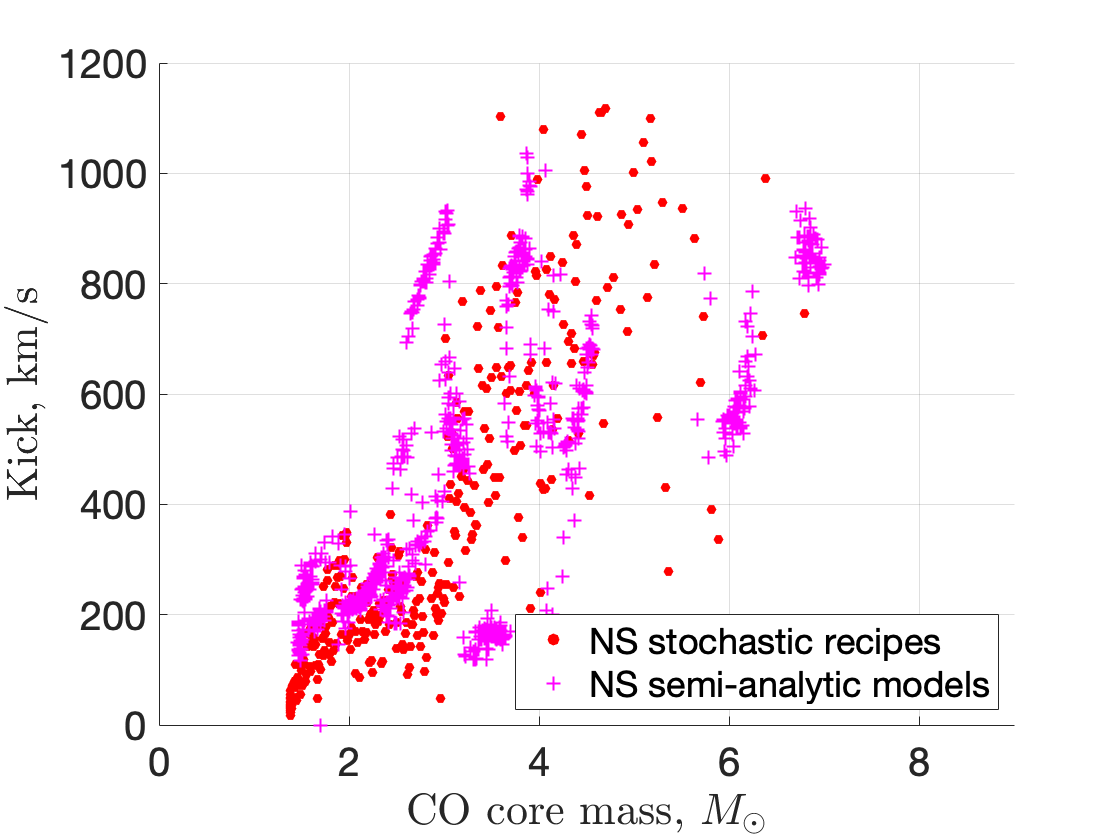}
   \caption{Neutron star natal kicks (3D velocities) as a function of CO core masses from single star simulations: a comparison between detailed stellar evolution coupled to the semi-analytic supernova model against COMPAS stellar evolution and our stochastic recipes.}
    \label{fig:kicks}
\end{figure}

In Figure \ref{fig:masses}, we compare the remnant masses as a function of progenitor CO core masses as predicted by a combination of the detailed stellar evolution and
the semi-analytic supernova model described in Section \ref{sec:models} against the probabilistic parametrised recipes introduced in Section \ref{sec:recipes}.  We implemented the stochastic recipes in the COMPAS rapid population synthesis code \citep{Stevenson:2017,VignaGomez:2018}.    

The recipes generally reproduce the key features of the more detailedsemi-analytic supernova model: the probabilistic nature of the remnant as a \ac{NS} or \ac{BH} and the approximate branching ratio of the two outcomes as a function of CO core mass; the approximate fraction of black holes undergoing complete fallback; and the two strands of the \ac{NS} mass distribution.  

Some of the fine features of the detailed models are, of course, not reproduced by the simplified recipes.  In certain cases (e.g., the difference in remnant mass for the most massive black holes that undergo complete fallback), this is due to differences between the pre-supernova models of Kepler \citep{Weaver:1978} used for the detailed simulations vs.~the single-stellar evolution models of \citet{Hurley:2000} employed in COMPAS.  Other fine features of detailed models are generally sufficiently uncertain at the moment that attempting to reproduce them appears pre-mature.  

For example, the detailed models show an island of consistent BH formation between CO core masses of $4.6 M_\odot$ and $6 M_\odot$, whereas the stochastic fit assumes a gradual increase
of the BH formation probability over a wider range in $\MCO$. This island can be traced to a peak in the core compactness of
the progenitor models that is associated
with the transition from convective to radiative
carbon burning \citep{Sukhbold:2018}.
While this peak appears quite robust
in single-star models, its location
and width are, however, sensitive to details
like the neutrino cooling rates
\citep{Mueller:2016}, the 
carbon-to-oxygen ratio after
helium burning \citep{Schneider:2020}
(which may be affected by previous
mass transfer episodes), and conceivably to other uncertainties in stellar models.

Another
example concerns the presence or absence
of a gap between \ac{NS} and \ac{BH} masses. While the detailed simulations support a narrow mass gap between the most massive neutron stars ($\approx 1.99 M_\odot$) and the lightest black holes ($\approx 2.75 M_\odot$), this gap is not enforced in the stochastic recipes.
This is motivated by the known limitations
of our semi-analytic treatment of fallback
that likely results in an overly clumpy
distribution of black hole masses
as discussed in Section~\ref{sec:models}.
Hydrodynamic models of fallback suggest a
remnant mass distribution without a gap
\citep{Ertl:2020}.
If the theoretical models (especially
multi-D simulations) reach greater clarity
on the presence or absence of a gap, the recipes can be updated in the future.

We show the neutron star natal kicks as a function of progenitor CO core masses in figure \ref{fig:kicks}.  The kicks do not reproduce the semi-analytic models very closely, particularly at the lowest CO core masses, where the recipe scaling of the kick with the ratio of the CO ejecta mass to the remnant mass, equation (\ref{eq:NSkick}), under-predicts the kicks relative to detailed simulations and asymptotes to zero for the least massive
cores. As argued previously, this is justified by
the premise behind our stochastic recipes: The overall
trends in the kick velocity predicted by the semi-analytic
model are likely robust, whereas the fine-grained structure
in the predicted distribution likely depends on more
uncertain details in the underlying stellar evolution
and explosion models.
The different asymptotic behaviour at low core masses
is intentional, however,
since it is line with the results of more sophisticated multi-D
simulations that cannot be easily reproduced in the semi-analytic
model. For the lowest CO core masses, the explosion develops
too quickly for significant global asymmetries to grow,
i.e.\ the assumption of a universal asymmetry parameter breaks down, and kicks may hence be as low as a few $\mathrm{km}\, \mathrm{s}^{-1}$. The lower kick values allowed by
the simplified recipes are also motivated by indications of reduced supernova kicks from low iron-core mass progenitors \citep{Podsiadlowski:2004}.

\begin{figure}
\centering
\includegraphics[width=\columnwidth]{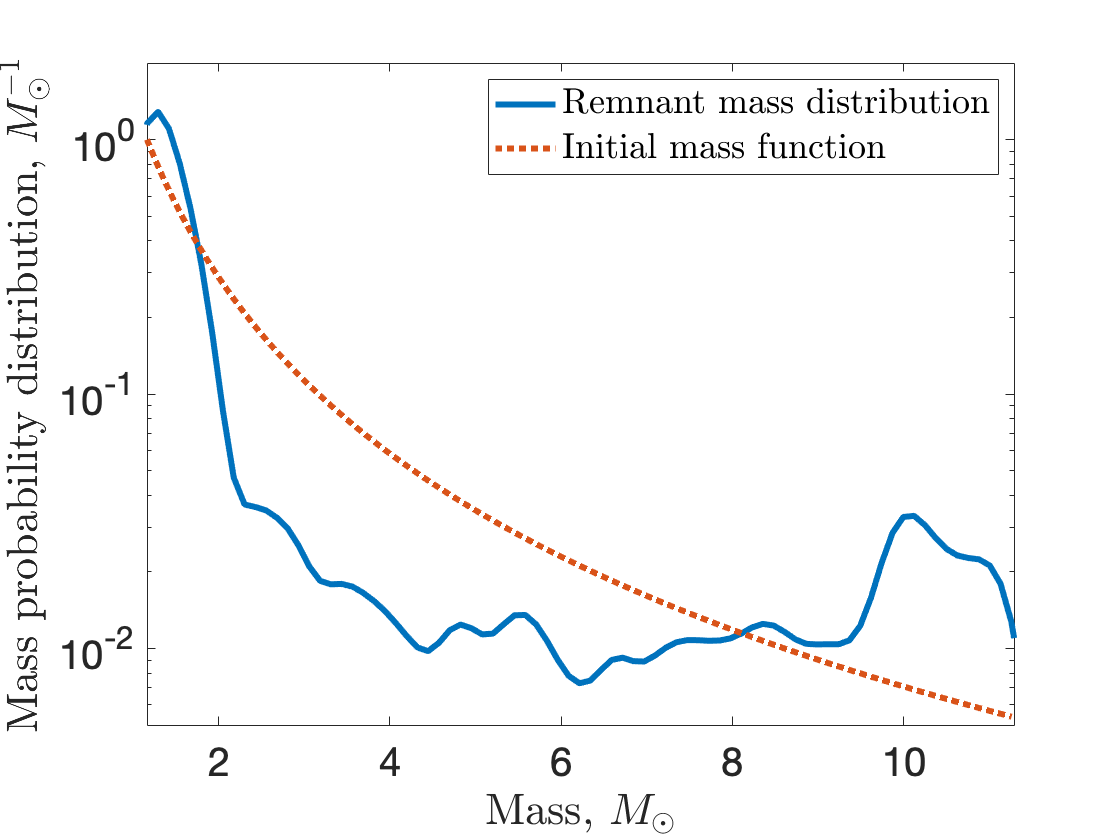}
   \caption{Remnant mass probability distribution (blue) for stars sampled from the initial mass function (dashed red).}
    \label{fig:massIMF}
\end{figure}

\begin{figure}
\centering
\includegraphics[width=\columnwidth]{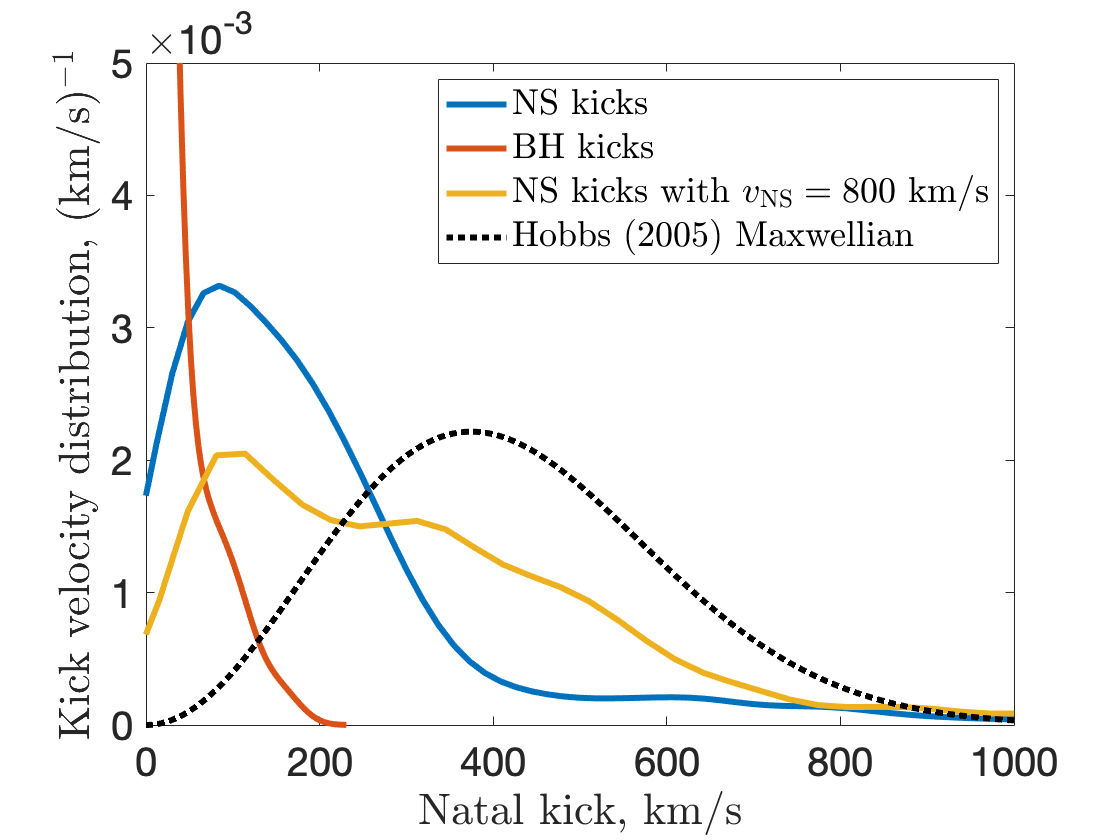}
   \caption{Natal kick probability distribution for neutron stars (blue) and black holes (red) for stars sampled from the initial mass function.  Neutron star kick distribution assuming $v_\mathrm{NS}=800$ km s$^{-1}$ is shown in yellow, while a Maxwellian distribution with a central parameter of 265 km s$^{-1}$ \citep{Hobbs:2005} is displayed in dashed black.}
    \label{fig:kickIMF}
\end{figure}

We can now investigate the predicted distributions of remnant masses and natal kicks in a population of single stars by integrating over the initial mass function.  We assume solar metallicity ($Z=0.0142$, \citealt{Asplund:2009}) and the standard COMPAS wind prescription for pre-supernova evolution.  We evolve stars with zero-age main sequence masses from $6 M_\odot$ (lower masses always leave CO white dwarfs behind) to $36 M_\odot$ (higher masses always lead to a complete collapse into black holes) following the \citet{Kroupa:2001} initial mass function, $p(M) \propto M^{-2.3}$ in this mass range.  The resulting remnant mass distribution is plotted in figure \ref{fig:massIMF}, along with the Kroupa initial mass function to guide the eye.

The corresponding kick distribution, separated into \ac{NS} and \ac{BH} kicks, is shown in figure \ref{fig:kickIMF}.  Two thirds of the initial-mass-function-weighted black holes for progenitors in this mass range receive zero natal kicks; the natal kicks of kicked \acp{BH} (shown in red) range up to $\sim 150$ km s$^{-1}$ with a root-mean-square kick of $\sim 60$ km s$^{-1}$.  

Neutron stars get larger kicks (blue), as expected, following a broad distribution extending from $\sim 15$ to $\sim 1500$ km s$^{-1}$, with a root-mean-square velocity of $\sim 270$ km s$^{-1}$ but a mode of only $\sim 80$ km s$^{-1}$.  This is lower than the Maxwellian distribution with a central parameter of 265 km s$^{-1}$ proposed by \citet{Hobbs:2005} (plotted in a black dotted curve in figure \ref{fig:kickIMF} for comparison).  However, the observed pulsar velocity distribution is uncertain, and a  number of other kick distributions have been fitted to radio pulsar observations \citep[e.g.,][]{Arzoumanian:2002,IgoshevVerbunt:2017}.  In particular, distance estimates based on dispersion measures are unreliable and could be systematically biased \citep{Deller:2019}, leading to errors in the inferred pulsar velocities.  Furthermore, there may be significant selection effects in the observed sample, ranging from potential correlations between velocity and detectability to only more rapidly kicked neutron stars escaping from binaries where many pulsars are born.  In any case, the parametrisation of our kick distribution allows for ready adjustment (e.g., the yellow line in figure \ref{fig:kickIMF} corresponds to $v_\mathrm{NS}=800$ km s$^{-1}$ rather than the default guess of $v_\mathrm{NS}=400$ km s$^{-1}$ ) and we leave observational constraints on the parameter values for a future investigation.  

Our model is consistent with the observation of large numbers of neutron stars in globular clusters.   These have a typical escape velocity of $\sim 50$ km s$^{-1}$, much smaller than most proposed kick velocity distributions, leading to the expectation that most neutron stars would be ejected at birth \citep[e.g.,][]{LyneLorimer:1994}.   If natal kicks are drawn from the \citet{Hobbs:2005} fit, fewer than $0.2$\% of all neutron stars would get kicks below $50$ km s$^{-1}$ and be retained in globular clusters. Meanwhile, the significant tail of low natal kicks in our model implies that 20\% of all neutron stars (and 30\% of all compact remnants) should be retained.  Even increasing $v_\mathrm{NS}$ to $800$ km s$^{-1}$, about $3\%$ of all neutron stars formed in globular clusters would remain there.  Here, we treat all stars as single and neglect the impact of binary companions on the retention fraction, either through increasing the effective inertial mass of a tight binary or through binary interactions that enhance the fraction of low-kick supernovae.

\section{Conclusion}\label{sec:conclusion}

We presented a set of recipes for compact-object remnant masses and natal kicks.  These recipes are derived from an examination of recent findings from
parametrised supernova models and first-principle 3D models.  They are probabilistic, accounting for the intrinsic stochasticity of stellar evolution and supernovae in the spread of outcomes (neutron stars or black holes) and the mass and kick values.  These recipes will be particularly useful for rapid stellar and binary population synthesis.  As an illustrative example, we showed the consequences of applying the recipes to the evolution of single massive stars.

Some of the key features of our prescriptions qualitatively match the observations.  For example, the absence of a mass gap between neutron stars and black holes\footnote{While this paper was under review, \citet{GW190814} announced the gravitational-wave discovery of a merging compact object binary, GW190814, whose lighter companion has a mass of $2.6 \pm 0.1$ M$_\odot$, placing in this putative mass gap.}  in the recipes is consistent with microlensing observations \cite{WyrzykowskiMandel:2019}.  The apparent mass gap from the analysis of black hole X-ray binaries \citep{Ozel:2010,Farr:2011} could be due to the evolutionary history of such systems or observational biases \citep{Kreidberg:2012}.   The non-negligible tail of low-kick neutron stars may help explain the retention of neutron stars in globular clusters, a challenge for high-kick models given typical escape velocities of $\lesssim 50$ km s$^{-1}$. 

The recipes contain 19 free parameters, of which 7 (highlighted in bold in table \ref{table:params}) are likely to be particularly uncertain and important.   While the supernova models provide guidance on their values, they do not uniquely determine them.  These free parameters should be constrained through a consistent application of all available observational evidence, including the masses and kicks of Galactic neutron stars observed as radio pulsars; the masses and velocities of neutron-star and black-hole X-ray binaries, including Be X-ray binaries; the retention of neutron stars and black holes in star clusters and globular clusters; microlensing measurements; and compact-object merger rates and properties determined from gravitational waves, short gamma-ray bursts, kilonovae, and r-process nucleosynthesis observations.

\section*{Acknowledgements}
We thank Ryosuke Hirai, Jeff Riley, Heloise Stevance, Simon Stevenson, Alejandro Vigna-Gomez, Reinhold Willcox, and other members of COMPAS and BPASS teams for useful discussions and suggestions.   IM and BM are recipient of the Australian Research Council Future Fellowships FT190100574 and FT160100035, respectively.
 
\section*{Data Availability}
The data underlying this article will be shared on reasonable request to the authors.  Simulations in this paper made use of the COMPAS rapid population synthesis code which is freely available at \url{http://github.com/TeamCOMPAS/COMPAS}.

\bibliography{Mandel}{}
\bibliographystyle{mnras} 

\label{lastpage}
\end{document}